\documentclass[preprint,showpacs,preprintnumbers,amsmath,amssymb]{revtex4}
\usepackage{booktabs}
\usepackage{mathrsfs}
\usepackage{epsfig}
\usepackage{graphicx}% Include figure files
\usepackage{dcolumn}% Align table columns on decimal point
\usepackage{bm}% bold math
\usepackage{amsmath}
\usepackage{slashed}       % slashed p

\let\jnfont=\rm
\def\NPB#1,{{\jnfont Nucl.\ Phys.\ B }{\bf #1},}
\def\PLB#1,{{\jnfont Phys.\ Lett.\ B }{\bf #1},}
\def\EPJC#1,{{\jnfont Eur.\ Phys.\ Jour.\ C }{\bf #1},}
\def\PRD#1,{{\jnfont Phys.\ Rev.\ D }{\bf #1},}
\def\PRL#1,{{\jnfont Phys.\ Rev.\ Lett.\ }{\bf #1},}
\def\MPLA#1,{{\jnfont Mod.\ Phys.\ Lett.\ A }{\bf #1},}
\def\JPG#1,{{\jnfont J.\ Phys.\ G}{\bf #1},}
\def\CTP#1,{{\jnfont Commun.\ Theor.\ Phys.\ }{\bf #1},}
\def\ZPC#1,{{\jnfont Z.\ Phys.\ C }{\bf #1},}
\def\JHEP#1,{{\jnfont JHEP \ }{\bf #1},}
\def\lsim{\raise0.3ex\hbox{$<$\kern-0.75em\raise-1.1ex\hbox{$\sim$}}}
\def\gsim{\raise0.3ex\hbox{$>$\kern-0.75em\raise-1.1ex\hbox{$\sim$}}}

\begin{document}

\title{Pair production of Higgs boson in NMSSM at the LHC
with next-to-lightest CP-even Higgs boson being SM-like}

\author{ Zhaoxia Heng, Xue Gong, Haijing Zhou}

\affiliation{
    College of Physics and Materials Science,
        Henan Normal University, Xinxiang 453007, China
      \vspace{1cm}}

\begin{abstract}
The next-to-minimal supersymmetric standard model (NMSSM) more naturally accommodates a Higgs boson with a mass of approximately 125 GeV than the minimal supersymmetric standard model (MSSM). In this work, we assume that the next-to-lightest CP-even Higgs boson $h_2$ is the SM-like Higgs boson $h$, whereas the lightest CP-even Higgs boson $h_1$ is dominantly singlet-like. We discuss the $h_1h_1$, $h_2h_2$, and $h_1h_2$ pair production processes via gluon-gluon fusion at the LHC for an collision energy of 14 TeV, and we consider the cases in which one Higgs boson decays to $b\bar b$ and the other one decays to $\gamma\gamma$ or $\tau^+\tau^-$. We find that, for $m_{h_1} \lesssim$ 62 GeV, the cross section of the $gg \to h_1 h_1$ process is relatively large and maximally reaches 5400 fb, and the production rate of the $h_1h_1\to b\bar b \tau^+\tau^-$ final state can reach 1500 fb, which make the detection of this final state possible for future searches of an integrated luminosity of 300 and 3000 $fb^{-1}$. This is mainly due to the contributions from the resonant production process $pp\to h_2\to h_1h_1$ and the relatively large branching ratio of $h_1\to b\bar b$ and $h_1\to\tau^+\tau^-$. The cross sections of the  $pp \to h_2h_2$ and $pp \to h_1 h_2$ production processes maximally reach 28 fb and 133 fb, respectively.
\end{abstract}


\maketitle

\section{Introduction}

In July 2012, the ATLAS and CMS collaborations at the CERN Large
Hadron Collider (LHC) \cite{1207ATLAS-CMS} reported the observation of a new
boson with a mass of approximately 125 GeV. After such discovery, which marked a milestone in particle physics, the extensive amount of data accumulated by both experiments \cite{LHC-experiment} have established that the properties of the new boson are consistent with those predicted by the standard model (SM). Since the discovery of the Higgs boson, massive work has been done to study its properties by measuring all the possible production channel cross sections and decay rates. In particular, it has been demonstrated that, among all the possible production modes, the double Higgs production is a sensitive probe for new physics \cite{hh-newphysics,hh-1617,hhyang,Ellwanger:2013ova,gghh-cao2013,hh-susy}.

At leading order in the SM, the Higgs pair production at the LHC happens through the gluon--gluon fusion $gg\to hh$ process \cite{Djouadi:1999rca,gghh}.
The SM cross section of such a process receives contributions from loops mediated by heavy quarks, which induce box and triangle diagrams ($gg\to h^*\to hh$). The amplitudes of these two types of diagrams interfere with each other destructively, thus leading to a rather small cross section. In new physics models, such as low energy supersymmetric models, both box and triangle diagrams receive additional contributions from the loops mediated by the
third generation of squarks. Moreover, the triangle diagrams are correlated with the trilinear Higgs coupling, which plays an important role in
the reconstruction of the Higgs potential \cite{SM-NLO-35fb}. Therefore, the $gg\to hh$ process is particularly sensitive to new physics contributions.

Among new physics models, the next-to-minimal supersymmetric standard model (NMSSM) has attracted considerable attentions \cite{Ellwanger:2009dp}, as it more naturally accommodates a Higgs boson with a mass near 125 GeV than the minimal supersymmetric standard model (MSSM). Compared with the MSSM, the NMSSM introduces an additional singlet Higgs field $\hat S$. Because of the coupling between singlet and doublet Higgs fields in the superpotential, the mass of the SM-like Higgs boson can be enhanced at the tree-level. Furthermore, the mixing between singlet and doublet Higgs fields can also increase the mass of the Higgs boson
when the next-to-lightest CP-even Higgs boson corresponds to the SM-like Higgs boson \cite{Cao:2012fz,Kang:2012sy,Cao-natural}. These two contributions can easily increase the mass of the Higgs boson to 125 GeV without any large radiative corrections. In this work, we only consider the scenario in which the next-to-lightest CP-even Higgs boson acts as the SM-like Higgs boson $h$.

Due to the presence of the additional superfield $\hat S$, the NMSSM contains three CP-even Higgs particles $h_i\ (i=1,2,3)$, which respect the convention
$m_{h_1}<m_{h_2}<m_{h_3}$. In this work, we assume that $h_2$ acts as the SM-like Higgs boson $h$, and $h_1$ is dominantly singlet-like considering the constraints for Higgs searches from LEP, Tevatron and LHC. We study the main di-Higgs production process at the LHC, the gluon--gluon fusion production of $h_1h_1$, $h_2h_2$, and $h_1h_2$ Higgs pairs. At present, the possible most promising detection channels for
Higgs pair production at the LHC are $hh\to b\bar b\gamma\gamma$, $hh\to b\bar b \tau^+\tau^-$ and $hh \to b\bar b W^+W^-$ \cite{hhsearch}. However, due to the weak interaction between the lightest CP-even Higgs $h_1$ and $W^+W^-$, we will focus on the $b\bar b\gamma\gamma$ and $b\bar b \tau^+\tau^-$ channels.

This work is organized as follows. In section II we briefly introduce the Higgs sector of the NMSSM. In section III we scan over the NMSSM parameter space
 by considering various experimental constraints; then we calculate the cross sections of $h_1h_1$, $h_2h_2$, and $h_1h_2$ di-Higgs gluon-fusion production processes and we also discuss the production rates for the $b\bar b\gamma\gamma$ and $b\bar b \tau^+\tau^-$ final states. Finally, we present our conclusions in section IV.

\section{Higgs sector in NMSSM}
As in the MSSM, the NMSSM contains two doublet Higgs superfields $\hat H_u$ and $\hat H_d$; however, in addition to the MSSM, the NMSSM introduces one gauge singlet superfield $\hat S$. As a consequence, the Higgs sector has a richer structure in the NMSSM than in the MSSM. To avoid the occurrence of dimensionful parameters in the superpotential, a discrete $Z_3$-symmetry is imposed. The corresponding superpotential and soft-breaking terms in the Higgs sector of the $Z_3$-invariant NMSSM are given by Equations (1) and (2) \cite{Ellwanger:2009dp,NMSSM}:
\begin{eqnarray} \label{NMSSM}
 W_{\rm NMSSM}&=&Y_u\hat{Q}\cdot\hat{H_u}\hat{U} -Y_d\hat{Q}\cdot\hat{H_d}\hat{D} -Y_e\hat{L}\cdot\hat{H_d}\hat{E}\nonumber\\
 &&+ \lambda\hat{H_u} \cdot \hat{H_d} \hat{S}
 + \frac{1}{3}\kappa \hat{S^3},\\
V_{\rm soft}^{\rm NMSSM}&=&\tilde m_u^2|H_u|^2 + \tilde m_d^2|H_d|^2
+ \tilde m_S^2|S|^2 \nonumber\\
&& +(A_\lambda \lambda SH_u\cdot H_d
+\frac{A_\kappa}{3}\kappa S^3 + h.c.).
\end{eqnarray}
where $\hat{Q}$, $\hat{U}$, $\hat{D}$, $\hat{L}$, and $\hat{E}$ are chiral superfields with $Y_u, Y_d, Y_e$ being their Yukawa coupling coefficients; $\lambda$ and $\kappa$ denote dimensionless coupling coefficients; and $\tilde{m}_{u}$, $\tilde{m}_{d}$, $\tilde{m}_{S}$, $A_\lambda$, and $A_\kappa$ are the soft-breaking parameters. When the electroweak symmetry is broken, the parameters $\tilde{m}_{u}$, $\tilde{m}_{d}$, and $\tilde{m}_{S}$
are expressed in terms of $m_Z$, $\tan\beta\equiv v_u/v_d$, and $\mu\equiv \lambda v_s$ with $v_u$, $v_d$, and $v_s$ being the vacuum expectation values (VEV) of the Higgs fields $H_u$, $H_d$, and $S$.
Thus, at the tree level, the Higgs sector of the NMSSM can be represented by six independent parameters given by Equation (3):
\begin{eqnarray}\label{parameter}
\lambda, \quad \kappa, \quad \tan \beta, \quad \mu, \quad  M_A, \quad A_\kappa,
\end{eqnarray}
with $M_A^2= \frac{2 \mu (A_\lambda + \kappa v_s)}{\sin 2 \beta}$.

In the NMSSM, $H_u$, $H_d$, and $S$ are given by Equation (4):
\begin{eqnarray}
H_u = \left ( \begin{array}{c} H_u^+ \\
       v_u +\frac{ \phi_u + i \varphi_u}{\sqrt{2}}
        \end{array} \right),~~
H_d & =& \left ( \begin{array}{c}
             v_d + \frac{\phi_d + i \varphi_d}{\sqrt{2}}\\
             H_d^- \end{array} \right),~~
S  =  v_s + \frac{1}{\sqrt{2}} \left(\sigma + i \xi \right).
\end{eqnarray}
By diagonalizing the corresponding mass matrices, we get the Higgs mass eigenstates as in Equation (5):
\begin{eqnarray} \left( \begin{array}{c} h_1 \\
h_2 \\ h_3 \end{array} \right) = U^H \left( \begin{array}{c} \phi_u
\\ \phi_d\\ \sigma\end{array} \right),~ \left(\begin{array}{c} a_1\\
a_2\\ G^0 \end{array} \right) = U^A \left(\begin{array}{c} \varphi_u
\\ \varphi_d \\ \xi \end{array} \right),~ \left(\begin{array}{c} H^+
\\G^+ \end{array}  \right) = U^C \left(\begin{array}{c}H_u^+\\ H_d^+
\end{array} \right).  \label{rotation}
\end{eqnarray}
Here, $h_i$ (with $i=1,2,3$) denotes the physical CP-even neutral scalars with $m_{h_1}<m_{h_2}<m_{h_3}$;  $a_i$ (with $i=1,2$) denotes the physical CP-odd neutral scalars with $m_{a_1}<m_{a_2}$; $H^+$ is the physical charged Higgs boson; and $G^0$ and $G^+$ are the Goldstone bosons absorbed by $Z$ and $W$ bosons, respectively.

Because of the coupling $\lambda\hat{H_u} \cdot \hat{H_d} \hat{S}$ in the superpotential, the mass of the SM-like Higgs boson $h$ at the tree level has the following dependencies:
\begin{equation}
 m^2_{h,tree}= m_Z^2\cos^22\beta+\lambda^2v^2\sin^22\beta.
\end{equation}
As shown by Equation (6), the SM-like Higgs boson ($h$) mass increases as $\lambda$ increases and $\tan\beta$ decreases. Moreover, the mixing between singlet and doublet Higgs fields can also significantly alter the $h$ mass. In particular, the mixing will pull down the mass when $h_1$ acts as $h$, whereas it will push up the mass when $h_2$ acts as $h$ \cite{Cao:2012fz,Kang:2012sy}. Therefore, when $h_2$ plays the role of the SM-like Higgs boson, both the additional tree-level contributions and the effect of Higgs doublet-singlet mixing can increase the $h$ mass; such phenomenon makes the large radiative corrections from top-squark loops unnecessary
to predict a Higgs boson with a mass near 125 GeV.

\section{Calculations and numerical results}

\subsection{Scanning strategy of the parameter space}

We perform a comprehensive scan over the parameter space of the NMSSM by using the package NMSSMTools \cite{NMSSMTools}. We constrain the mass of the gluino and the soft breaking parameters in the first two generations of the squark sector to be equal to 2 TeV. Moreover, we assume a common value denoted by $m_{\tilde{l}}$ for all soft breaking parameters in the slepton sector. The parameter space considered in the scan is represented by Equation (7):
\begin{eqnarray}\label{NMSSM-scan}
&& 0 <\lambda,\kappa\leq 0.75, \quad  2 \leq \tan{\beta} \leq 60, 100{\rm ~GeV}\leq \mu \leq 1 {\rm ~TeV},
     \quad 50 {\rm ~GeV}\leq M_A \leq 2 {\rm ~TeV}, \nonumber \\
&&  |A_{\kappa}| \leq 2 {\rm TeV},
\quad  100{\rm ~GeV}\leq m_{\tilde{l}} \leq 1 {\rm ~TeV}, 100{\rm ~GeV}\leq M_{Q_3}\leq 2 {\rm ~TeV}, \nonumber\\
&& 100{\rm ~GeV}\leq M_{U_3}(=M_{D_3}) \leq 2 {\rm ~TeV}, |A_{t}(=A_{b})|\leq {\rm min}(3 \sqrt{M_{Q_3}^2 + M_{U_3}^2}, 5 {\rm TeV}), \nonumber\\
&& 20 {\rm GeV} \leq M_1 \leq 500 {\rm GeV},100 {\rm GeV} \leq M_2 \leq 1 {\rm TeV}.
\end{eqnarray}
In addition to the constraints implemented in the package NMSSMTools we also consider the following:
\begin{itemize}
  \item Constraints from the direct searches for the Higgs boson at LEP, Tevatron, and LHC. We adopt the package HiggsSignals for the fit of the 125 GeV Higgs data \cite{HiggsSignal} and the
package HiggsBounds for the search of non-standard Higgs bosons at colliders  \cite{HiggsBounds}.
  \item Constraints from direct searches of sparticles at the LHC RunI and Run-II. We use the packages FastLim \cite{FastLim} and SModelS \cite{SModelS} to implement such constraints. In particular, the results of the experimental searches for gluino and squarks are implemented in the package FastLim, whereas those of sleptons and electroweakinos searches are implemented in the SModelS package. Efficiencies or upper bounds on the sparticle productions can be obtained to limit the parameter space of the NMSSM.
  \item Constraints from the searches for electroweakinos and top-squarks at the LHC  Run-II. The surviving samples that satisfy the constraints from
FastLim and SModelS are further tested by detailed Monte Carlo simulations.
In details, we use MadGraph/MadEvent \cite{madgraph} to simulate events at the parton level and Pythia \cite{pythia} to do parton shower and hadronization, then we put the events into Delphes \cite{delphes} to do detecor simulation. Finally, we implement the ATLAS and CMS searches within CheckMATE \cite{checkmate} to decide whether each parameter point in our scenaio is allowed or not.
\end{itemize}

\subsection{Numerical results}
In our numerical calculation we take $m_t$ = 173.2GeV, $m_b$ = 4.18GeV,
 $m_Z$ = 91.19GeV \cite{PDG} and take the renormalization and factorization scales to be the invariant mass of the Higgs pair. We
 select the samples for 122 GeV $\leq m_h \leq$ 128 GeV
and require the next-to-lightest CP-even Higgs boson ($h_2$) to be
the SM-like Higgs boson $h$.

%%Fig1 %%%%%%%%%%%%%%%%%%%%%%%%%%%%%%%%%%%%%%%%%%%%%%%%%%%%%%%%%%%%%%%%%%
\begin{figure}[t]
\includegraphics[width=15cm]{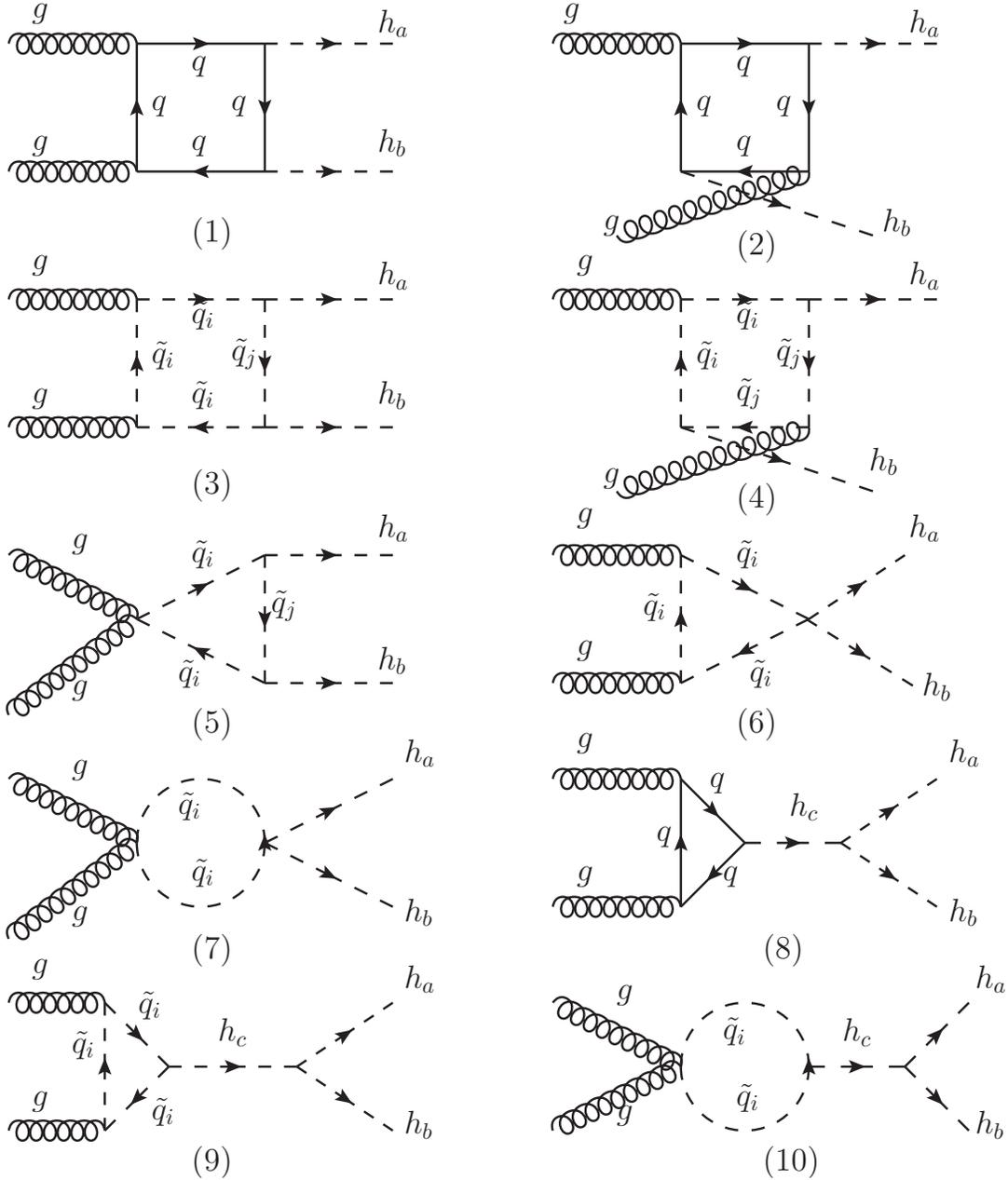}
\vspace{-0.5cm}
\caption{Feynman diagrams of Higgs pair production in the NMSSM
with a,b=1,2 and c=1,2,3.}
\label{fig1}
\end{figure}
%%%%%%%%%%%%%%%%%%%%%%%%%%%%%%%%%%%%%%%%%%%%%%%%%%%%%%%%%%%%%%%%%%%%%%%%%

According to the NMSSM scenario, the Higgs pair production processes possibly available at the LHC happen via
gluon--gluon fusion, as shown in Fig. \ref{fig1}. Differently from the SM, in the NMSSM the gluon--gluon fusion receives additional contributions from the loops mediated by the third generation of squarks. We checked that we can reproduce the results in the SM and MSSM presented in \cite{Djouadi:1999rca} and \cite{gghh-MSSM}.
In the numerical calculations, for an LHC collision energy of 14 TeV, we find that the cross section of the $gg\to hh$ process in the SM scenario is 18.7 fb at $m_h =$ 125 GeV.

\begin{figure}[t]
\includegraphics[width=6.3cm]{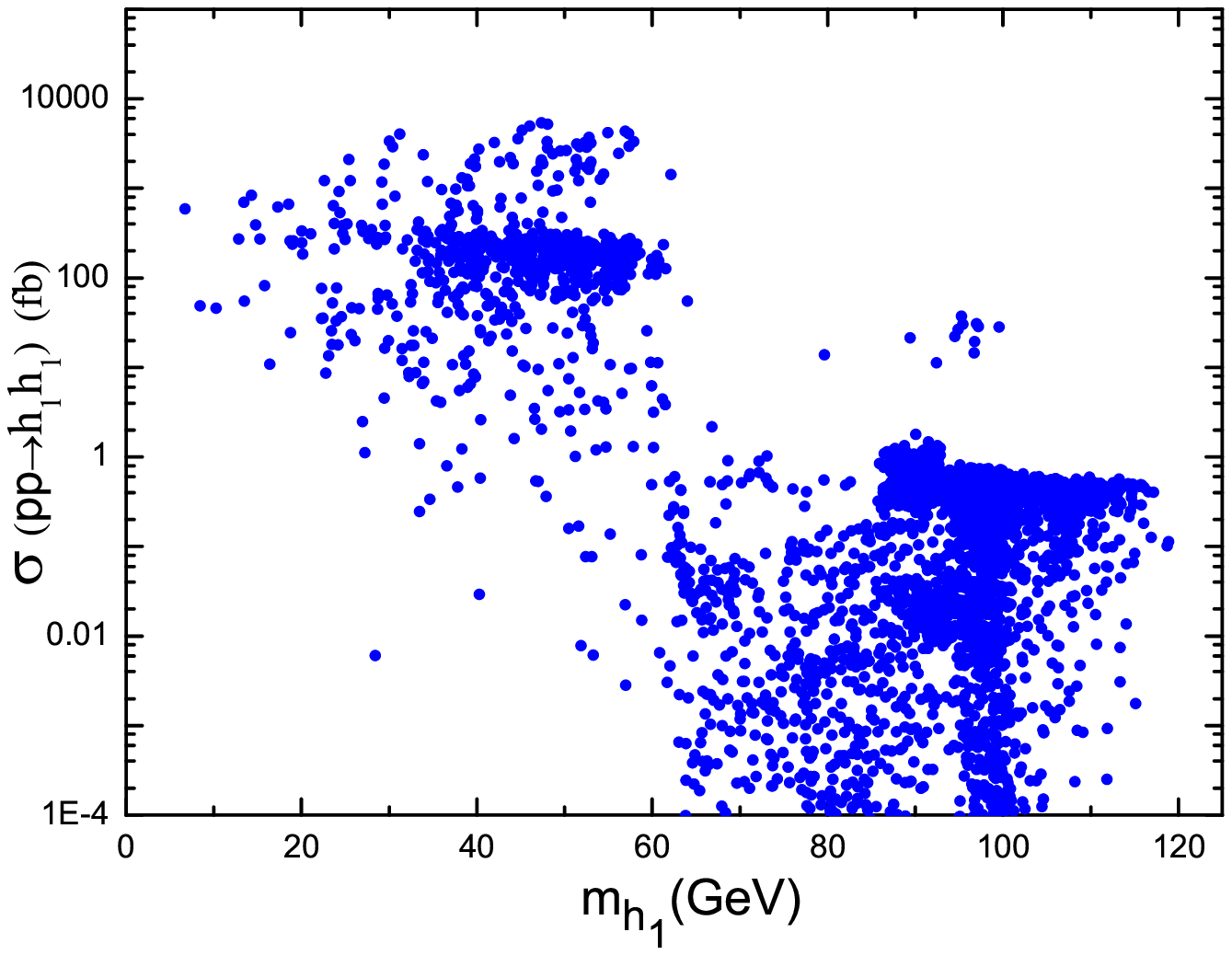}\hspace{-0.8cm}
\includegraphics[width=6.3cm]{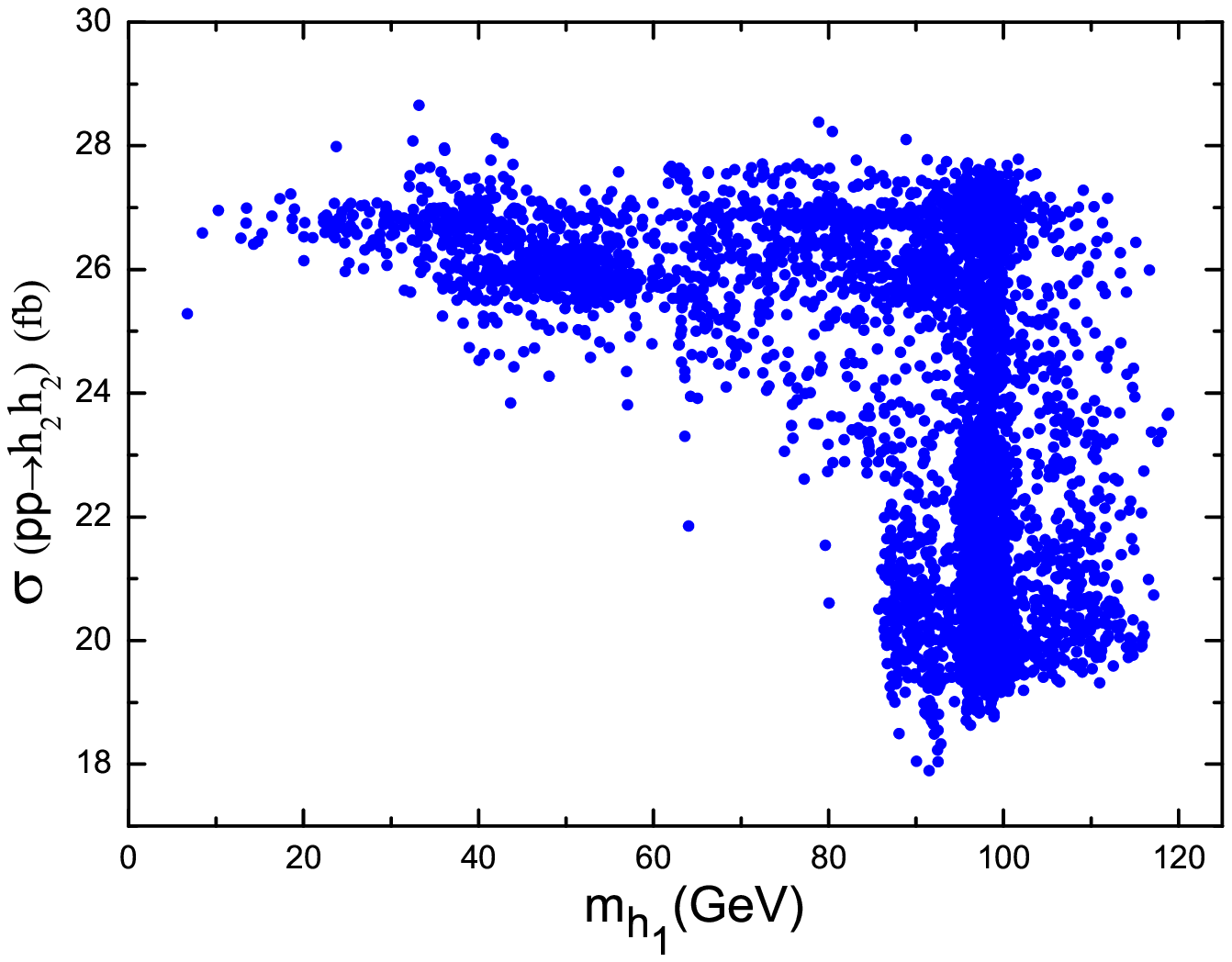}\hspace{-0.8cm}
\includegraphics[width=6.3cm]{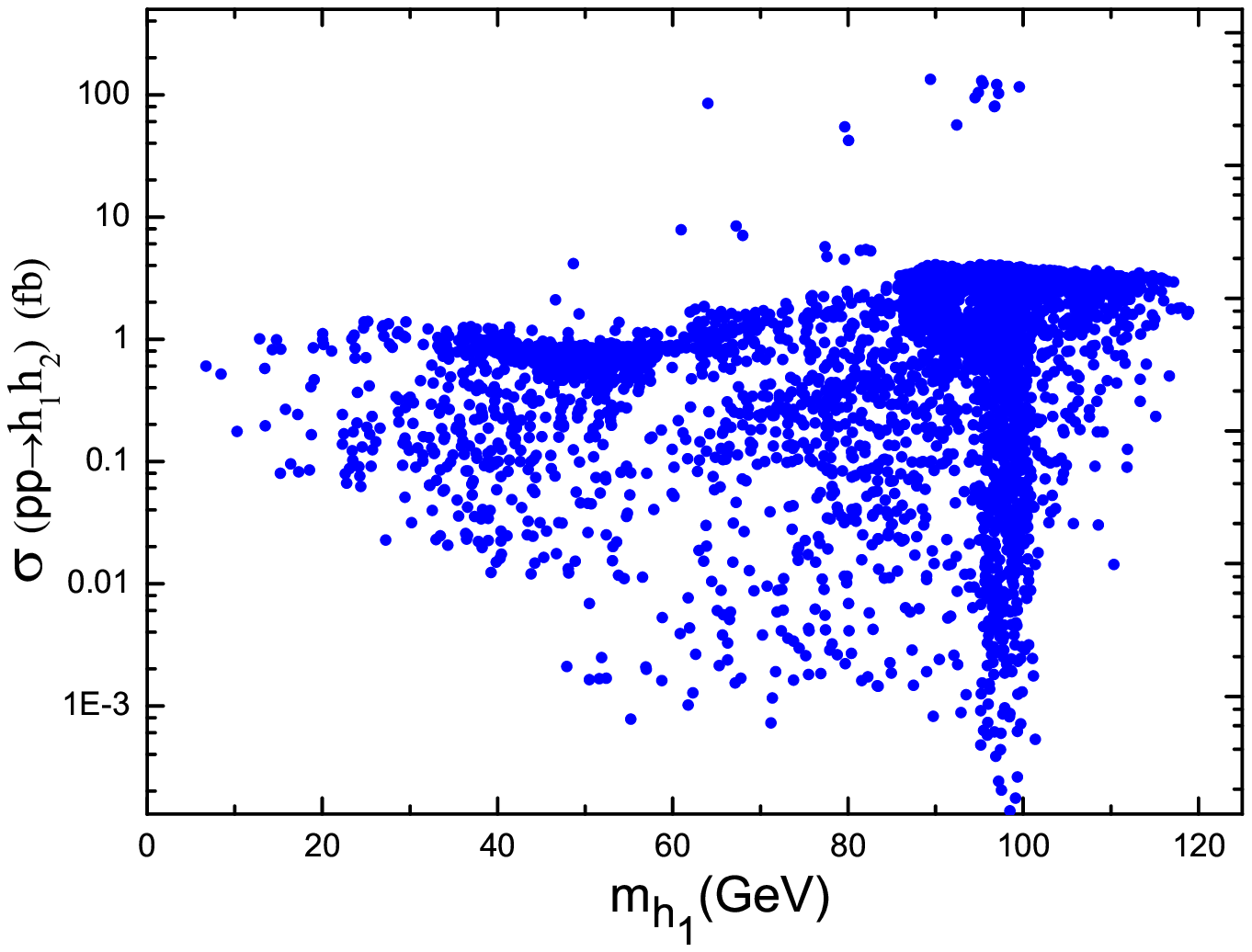}
\vspace{-0.2cm}
\caption{Cross sections of Higgs pair production processes in the final states $h_1h_1$, $h_2h_2$, and $h_1h_2$.}
\label{fig2}
\end{figure}

\begin{figure}
\includegraphics[width=9cm]{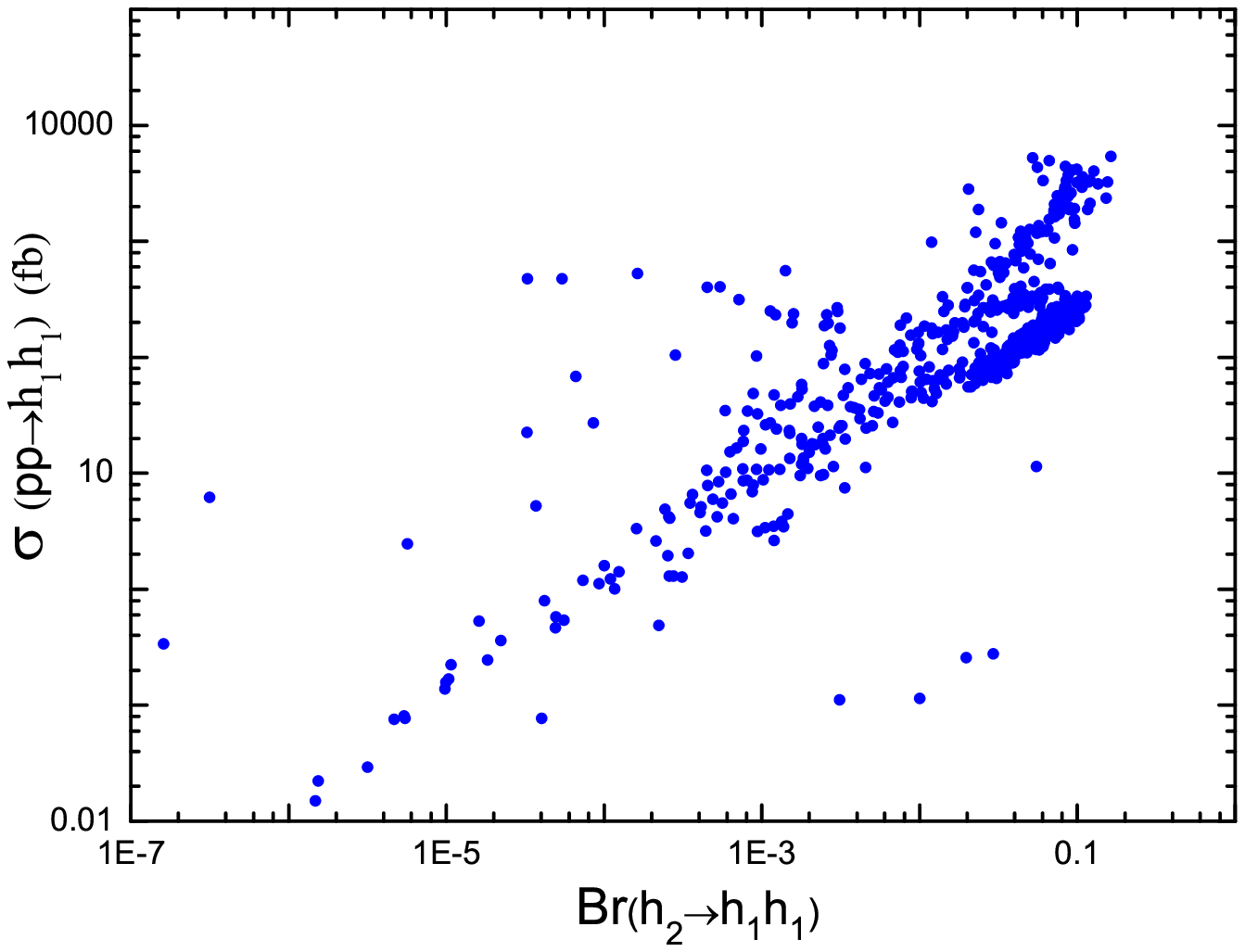}
\caption{Cross section of the Higgs pair production process in the final state $h_1h_1$ as a function of the $h_2\to h_1h_1$ branching ratio.}
\label{fig3}
\end{figure}

Figure \ref{fig2} shows the cross sections of the $h_1h_1$ (left), $h_2h_2$ (middle), and $h_1h_2$ (right) pair production processes. The cross section of the process $pp \to h_1 h_1$ (left panel) assumes values mainly divided into two regions: for $m_{h_1} \lesssim$ 62 GeV, the cross section is relatively large and it can maximally reach 5400 fb. This is because the rare decay $h_2\to h_1h_1$ is open when $m_{h_1} \lesssim$ 62 GeV. As can be clearly seen from Fig. \ref{fig3}, which shows the cross section of the $pp\to h_1h_1$ process as a function of the branching ratio of the decay $h_2\to h_1h_1$, the $pp\to h_1h_1$ cross section increases as the $h_2\to h_1h_1$ branching ratio increases.

%%Fig2 %%%%%%%%%%%%%%%%%%%%%%%%%%%%%%%%%%%%%%%%%%%%%%%%%%%%%%%%%%%%%%%%%%
%%%%%%%%%%%%%%%%%%%%%%%%%%%%%%%%%%%%%%%%%%%%%%%%%%%%%%%%%%%%%%%%%%%%%%%%%
%%Fig3 %%%%%%%%%%%%%%%%%%%%%%%%%%%%%%%%%%%%%%%%%%%%%%%%%%%%%%%%%%%%%%%%%%
%%%%%%%%%%%%%%%%%%%%%%%%%%%%%%%%%%%%%%%%%%%%%%%%%%%%%%%%%%%%%%%%%%%%%%%%%

The plot in the middle panel of Fig. \ref{fig2} indicates that the cross section of the $pp\to h_2h_2$ process can maximally reach 28 fb, which is 1.5 times larger than the SM prediction. The enhancement mainly originates from the contributions of the diagrams (3), (4), and (5) in Fig. \ref{fig1}. For lighter $m_{\tilde t_1}$ and larger values of $A_t$ the cross section is larger, which is in agreement with the results reported in \cite{gghh-cao2013}.

From the right panel of Fig. \ref{fig2} it is visible that, for the majority of the surviving samples, the cross section of $pp\to h_1h_2$ is less than 10 fb; this mainly because $h_1$ is dominantly singlet-like. However, for a few surviving samples, the cross section can exceed 100 fb and reach 133 fb at maximum. In these cases, the heaviest CP-even Higgs $h_3$ is relatively light with a mass of approximately 500 GeV, and $h_3$ is produced on-shell with the branching ratio to $h_1h_2$ being about 10\%.
It is important to notice that the mass of the heavy Higgs boson near 500 GeV for $\tan\beta\lesssim 5$ in the MSSM is not excluded by experimental results based on the LHC Run-II data \cite{heavy Higgs-ATLAS, heavy Higgs-CMS}.

\begin{figure}[t]
\includegraphics[width=8.5cm]{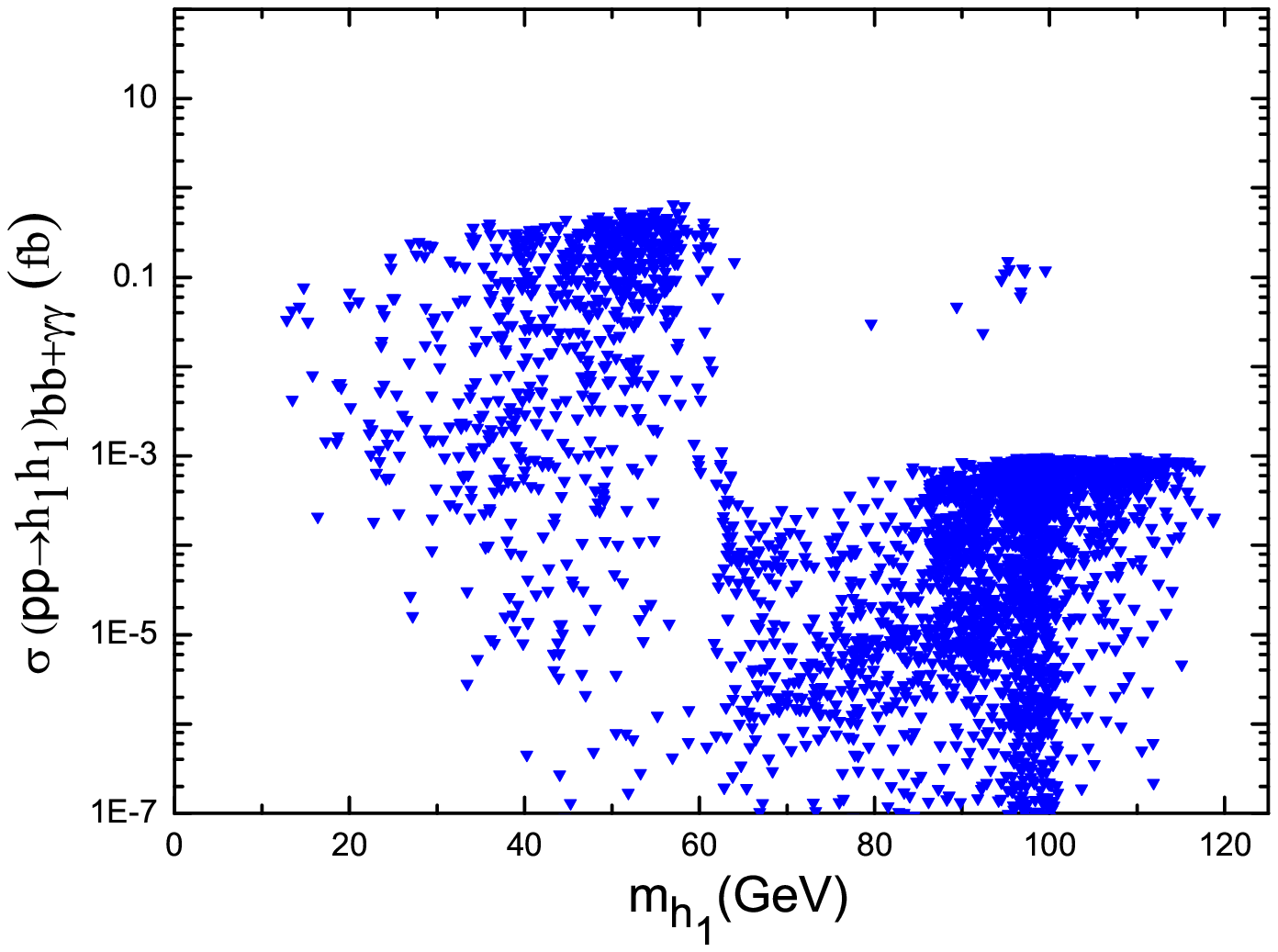}\hspace{-1.5cm}
\includegraphics[width=8.5cm]{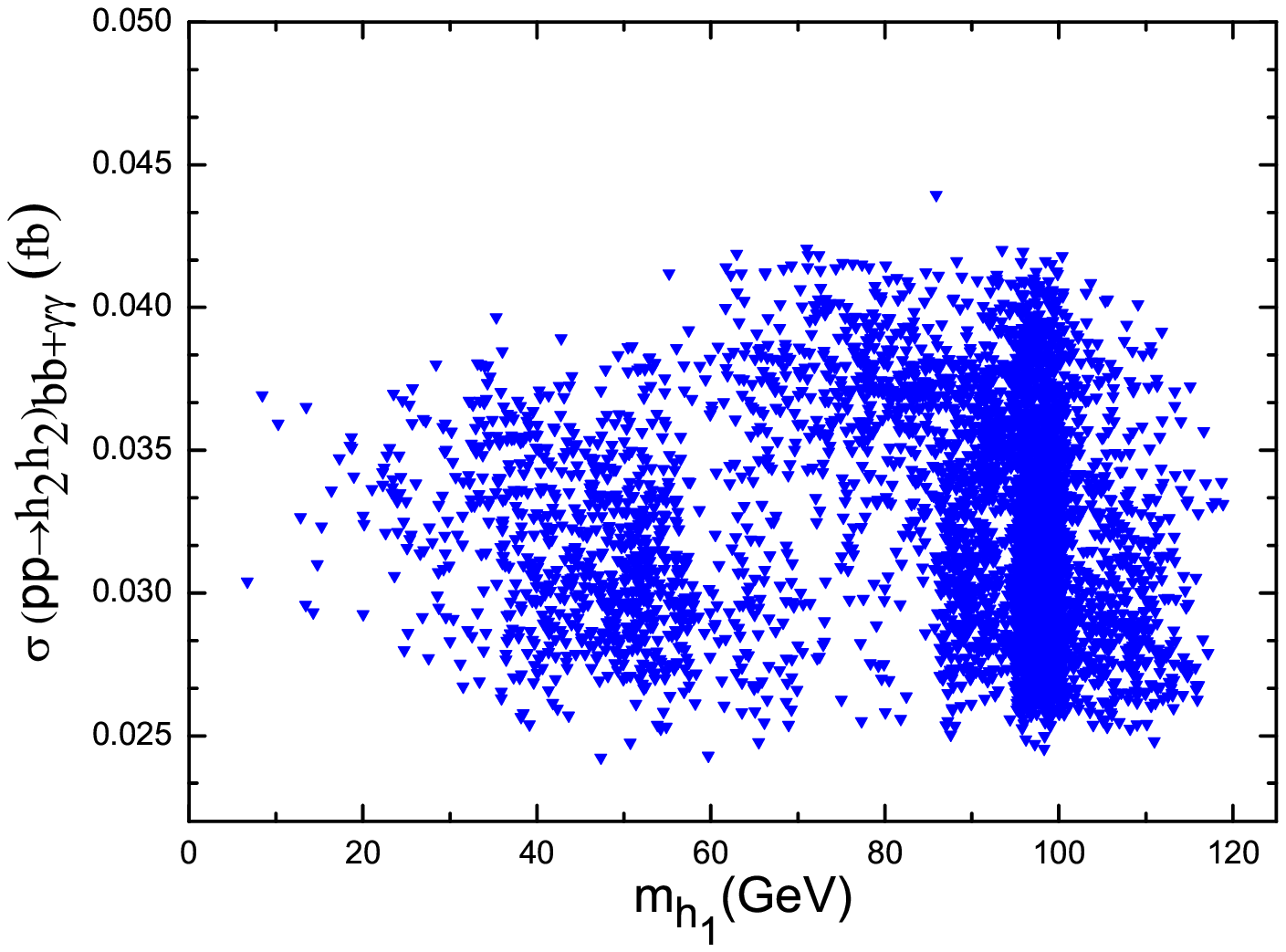}
\includegraphics[width=8.5cm]{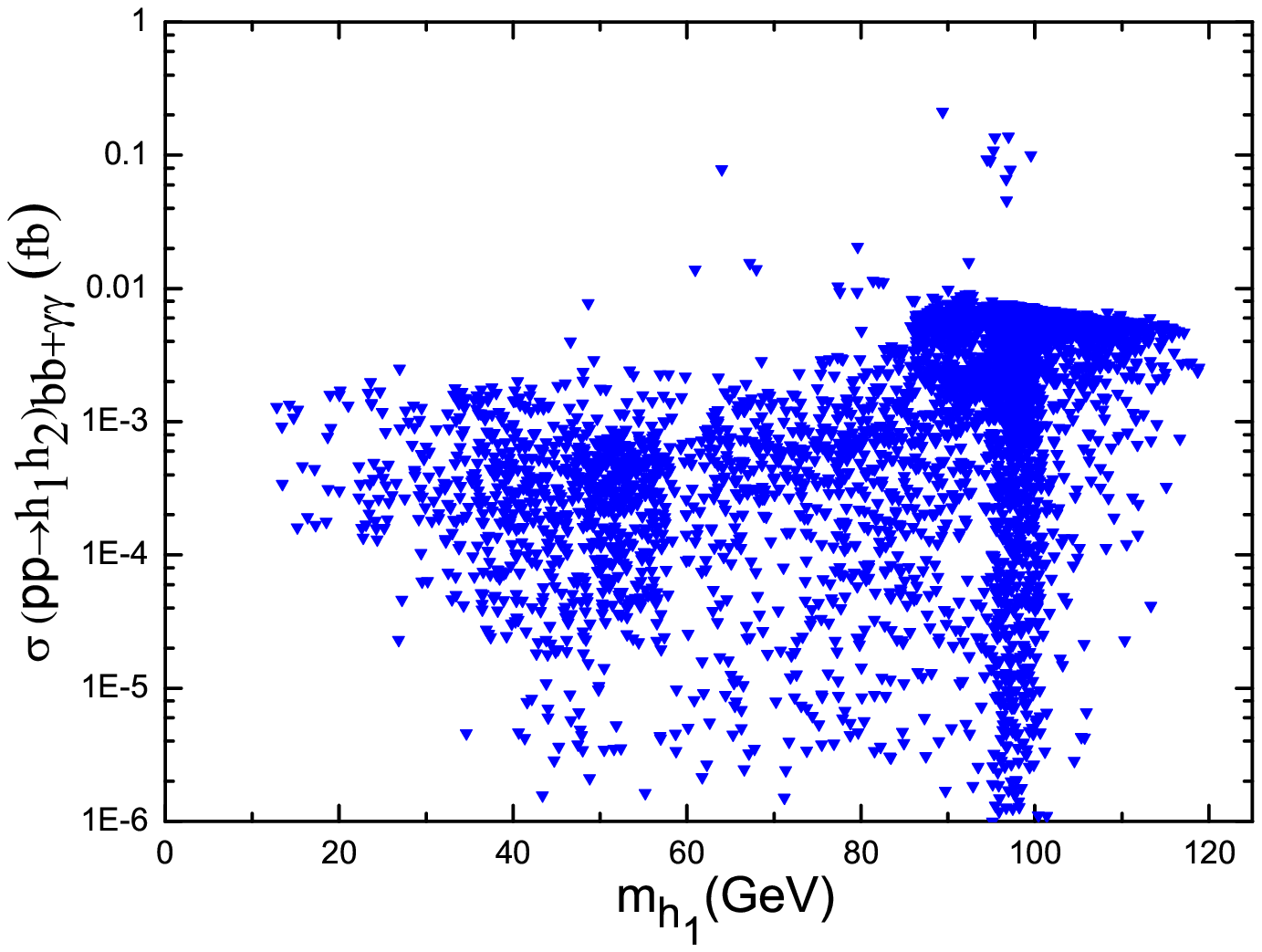}\hspace{-1.5cm}
\includegraphics[width=8.5cm]{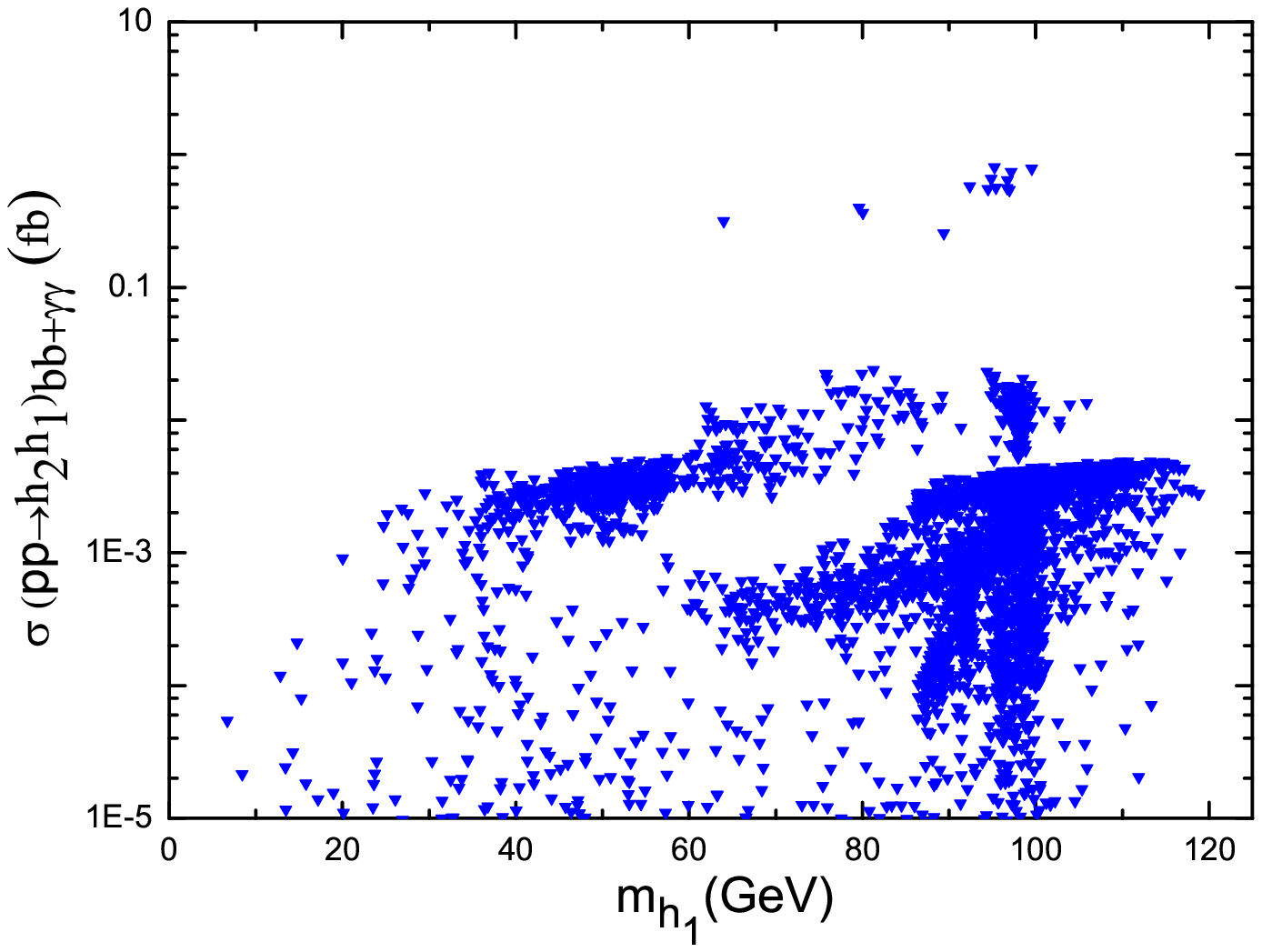}
\vspace{-0.2cm}
\caption{Production rates of Higgs pair production processes in the final states $h_1h_1\to b\bar b \gamma\gamma$, $h_2h_2\to b\bar b \gamma\gamma$, $h_1h_2\to b\bar b \gamma\gamma$, and $h_2h_1\to b\bar b \gamma\gamma$.}
\label{fig4}
\end{figure}

\begin{figure}[t]
\includegraphics[width=8.5cm]{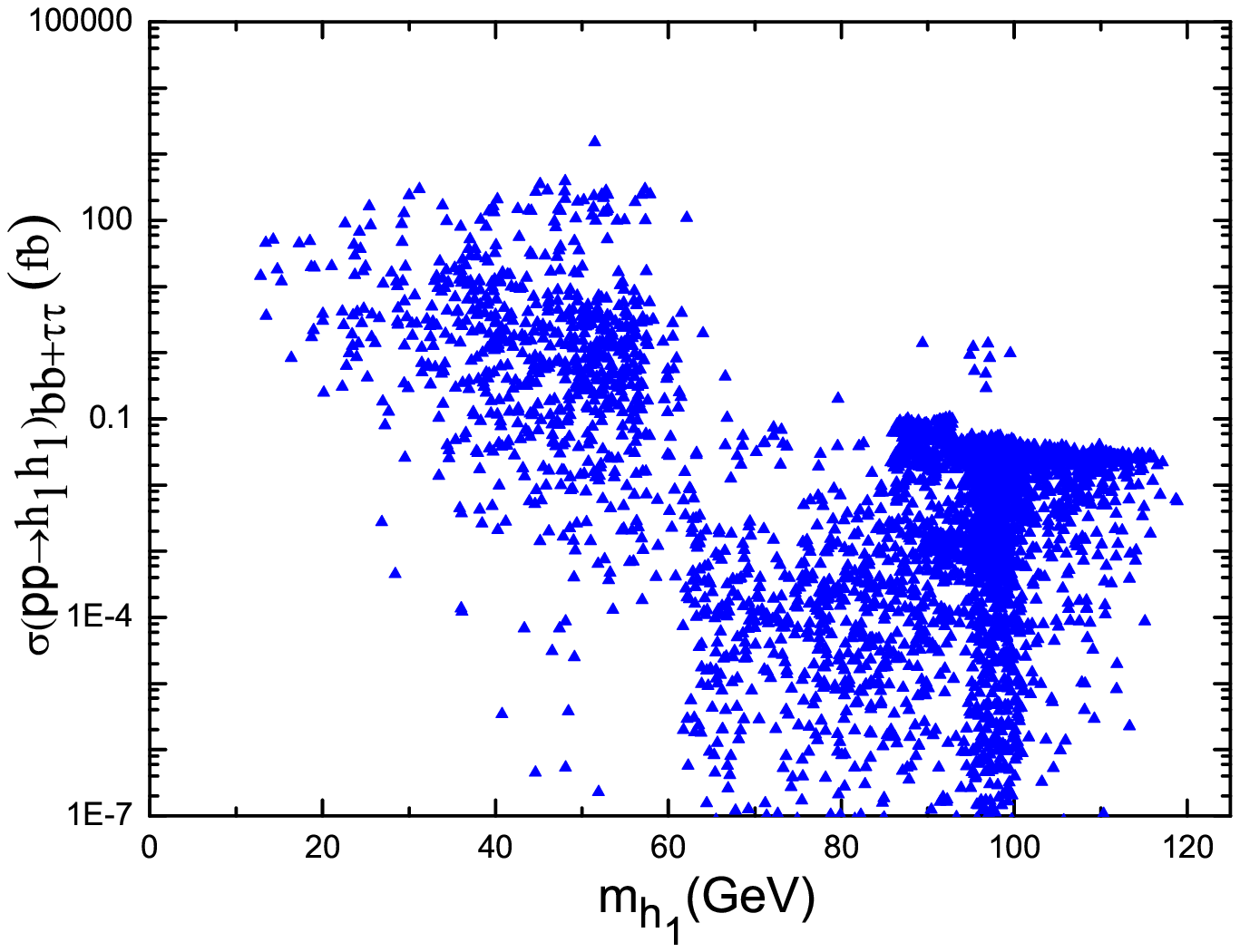}\hspace{-1.5cm}
\includegraphics[width=8.5cm]{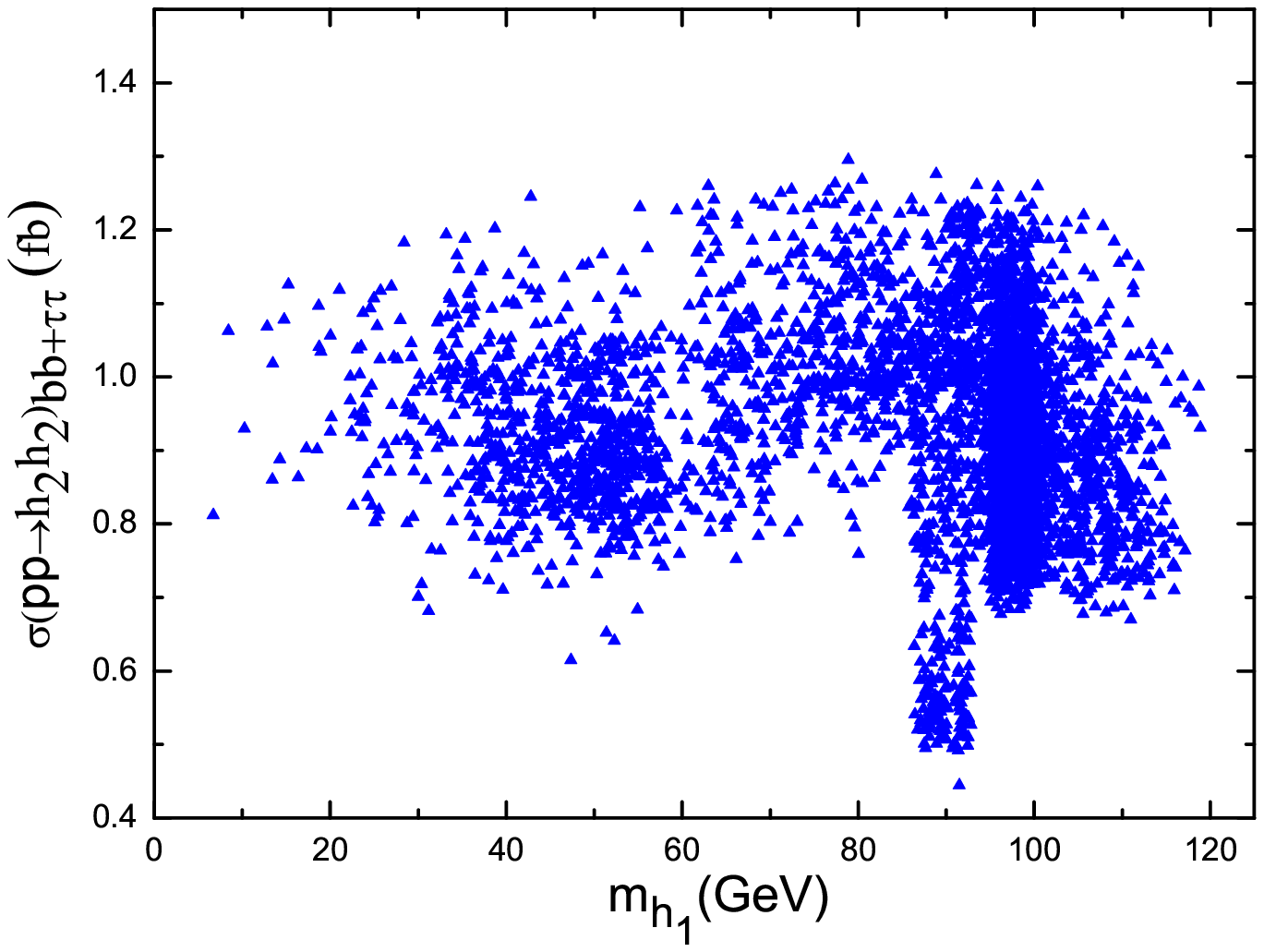}\vspace{-0.8cm}
\includegraphics[width=8.5cm]{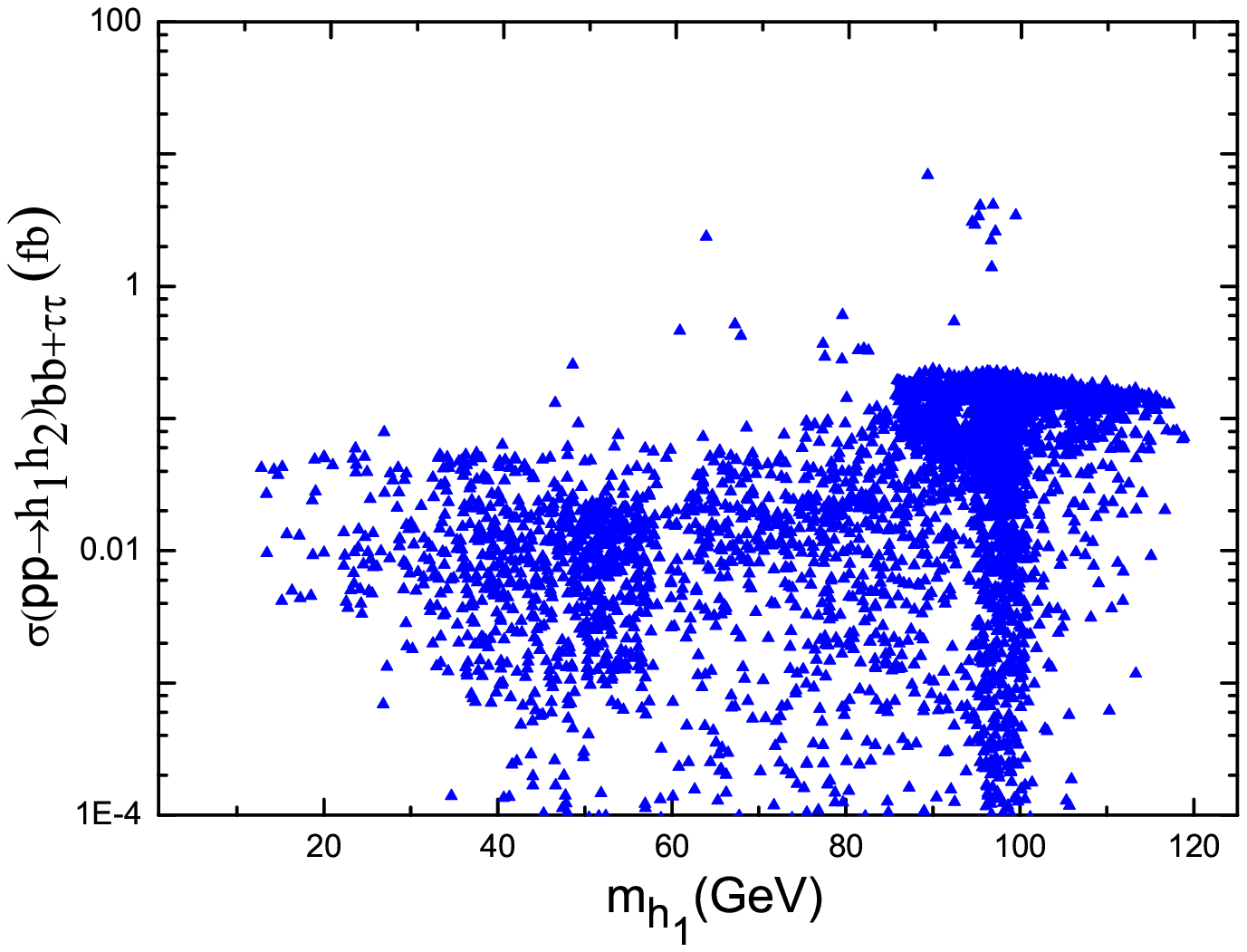}\hspace{-1.5cm}
\includegraphics[width=8.5cm]{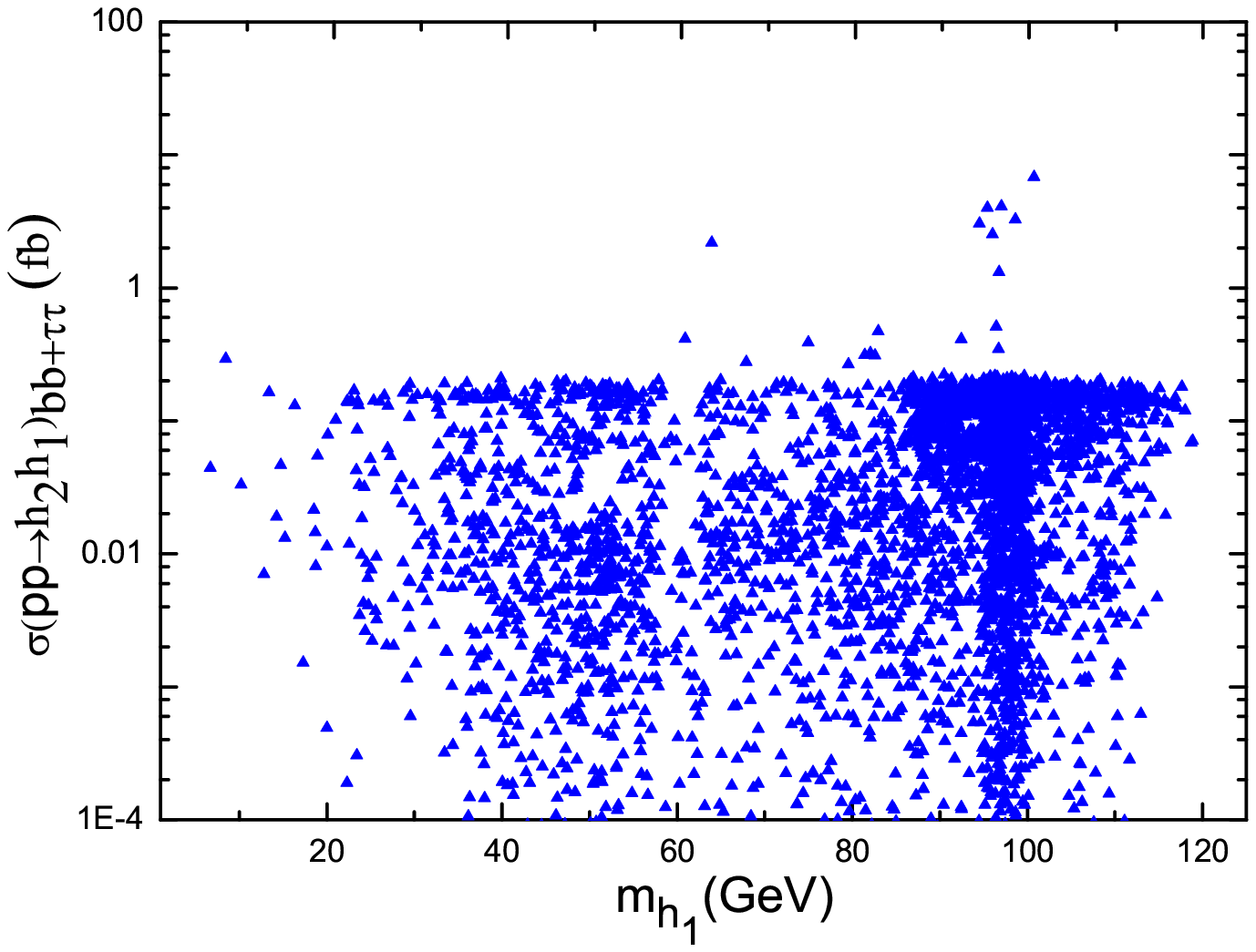}
\vspace{-0.2cm}
\caption{Production rates of Higgs pair production processes in the final states $h_1h_1\to b\bar b \tau^+\tau^-$, $h_2h_2\to b\bar b \tau^+\tau^-$, $h_1h_2\to b\bar b \tau^+\tau^-$, and $h_2h_1\to b\bar b \tau^+\tau^-$.}
\label{fig5}
\end{figure}

Since the discovery potential of the Higgs boson depends on the decay spectrum of the Higgs boson, we now consider the $h_1h_1$, $h_2h_2$, and $h_1h_2$ di-Higgs production processes with one Higgs boson decaying to $b\bar b$ and the other one decaying to either $\gamma\gamma$ or $\tau^+\tau^-$.
In Fig. \ref{fig4} and Fig. \ref{fig5} we show the production rates of the final states $b\bar b \gamma\gamma$ and $b\bar b \tau^+\tau^-$, respectively.
As it can be seen, the production rate of the process $h_1h_1\to b\bar b \tau^+\tau^-$ is relatively large for $m_{h_1}\lesssim$ 62 GeV, and it can maximally reach 1500 fb. This is because, for most of the surviving samples, the $h_1\to b\bar b$ and $h_1\to\tau^+\tau^-$ branching ratios are close to 90\% and 8\%, respectively.

%%Fig4 %%%%%%%%%%%%%%%%%%%%%%%%%%%%%%%%%%%%%%%%%%%%%%%%%%%%%%%%%%%%%%%%%%
%%%%%%%%%%%%%%%%%%%%%%%%%%%%%%%%%%%%%%%%%%%%%%%%%%%%%%%%%%%%%%%%%%%%%%%%%
%%Fig5 %%%%%%%%%%%%%%%%%%%%%%%%%%%%%%%%%%%%%%%%%%%%%%%%%%%%%%%%%%%%%%%%%%
%%%%%%%%%%%%%%%%%%%%%%%%%%%%%%%%%%%%%%%%%%%%%%%%%%%%%%%%%%%%%%%%%%%%%%%%%

Both the ATLAS and CMS collaborations performed searches for 125GeV Higgs pair production in the final states $b\bar b b\bar b$, $b\bar b\tau^+\tau^-$ and $b\bar b \gamma\gamma$.  Based on the data collected at
$\sqrt{s}$ = 13 TeV with an integrated luminosity of 35.9 $fb^{-1}$,
the observed 95\% CL upper limit for $b\bar b b\bar b$ final state is 147fb
corresponding to 13 times the SM prediction \cite{bbbb-atlas}, for $b\bar b\tau^+\tau^-$ final state is 75.4fb corresponding to 30 times the SM prediction \cite{bbtau-CMS} and for $b\bar b \gamma\gamma$ final state is 1.67fb corresponding to 19.2 times the SM prediction \cite{bbgamma-CMS}.
For the present integrated luminosity of LHC, the Higgs pair production in the considered scenario of NMSSM cannot be directly detected. Due to the large production rate of the process $pp\to h_1h_1\to b\bar b\tau^+\tau^-$, for future searches of an integrated luminosity of 300 and 3000 $fb^{-1}$, Higgs pair production for $h_1h_1\to b\bar b\tau^+\tau^-$ final state may be detected.

\section{Conclusion}
Because of the mixing effect of Higgs doublet-singlet fields, and the coupling
effect between Higgs doublet and singlet fields in the superpotential,
the NMSSM more naturally accommodates a Higgs boson with a mass of approximately
125 GeV. We assume that the next-to-lightest CP-even Higgs boson $h_2$ plays the role
of the SM-like Higgs boson. In this scenario the lightest CP-even
Higgs boson $h_1$ is dominantly singlet-like and it can be much lighter
than 125 GeV. In this work, we discussed the $h_1h_1$, $h_2h_2$, and $h_1h_2$ pair production processes via gluon-gluon fusion at the
LHC for an collision energy of 14 TeV. In addition, we studied the production rate of such di-Higgs events in which one Higgs boson
decays to $b\bar b$ and the other one to either $\gamma\gamma$ or $\tau^+\tau^-$. We found that, for $m_{h_1}\lesssim$ 62 GeV, the cross section of the $gg \to h_1h_1$ production process is relatively large and maximally reaches  5400 fb, and that the production rate of the $h_1h_1\to b\bar b \tau^+\tau^-$ final state can reach 1500 fb. These results are mainly due to the contributions from the resonant production process $pp\to h_2\to h_1h_1$and the relatively large branching ratio of the $h_1\to b\bar b$ and $h_1\to\tau^+\tau^-$ decays. The cross sections of the  $pp \to h_2h_2$ and $pp \to h_1 h_2$ production processes maximally reach 28 fb and 133 fb, respectively.
Therefore, for future searches of an integrated luminosity of 300 and 3000 $fb^{-1}$, Higgs pair production for $h_1h_1\to b\bar b\tau^+\tau^-$ final state may be detected.

\section*{Acknowledgement}
We thank Prof. Junjie Cao and Dr. Liangliang Shang for helpful discussions in the revision of the work. This work was supported in part by the National Natural Science Foundation of China (NNSFC) under grant No. 11305050 and 11705048.
%%%%%%%%%%%%%%%%%%%%%%%%%%%%%%%%%%%%%%%%%%%%%%%%%%%%%%%%%%%%%%%%%%%%%%%%%
%%%%%%%%%%%%%%%%%%%%%%%%%%%%%%%%%%%%%%%%%%%%%%%%%%%%%%%%%%%%%%%%%%%%%%%%%

\end{document}